
\magnification 1200
\baselineskip=14pt
\def\ni{\noindent}
\def\cel{\centerline}
\def\Par{\par \vskip .23cm}

\def\d{{\bf d}}
\def\w{{\bf\omega}}

\def\qleft{\char'134}
\def\qright{\char'42}
\baselineskip 14pt
\cel { \bf THERMODYNAMICS OF THE STEPHANI UNIVERSES$^*$}
\Par

\baselineskip 12pt

\cel { Hernando Quevedo and Roberto A. Sussman}
\cel { Instituto de Ciencias Nucleares, UNAM}
\cel { Apartado Postal 70-543,  M\'exico D.F.04510, \ M\'EXICO} \Par
\Par

\vskip .5 cm

\cel { ABSTRACT} \par

We examine the consistency of the thermodynamics of
the most general class of conformally flat solution with an irrotational
perfect fluid source (the Stephani Universes). For the case when the
isometry group has dimension $r\ge2$, the Gibbs-Duhem relation is always
integrable, but if $r<2$ it is only integrable for the particular subclass
(containing FRW cosmologies) characterized by $r=1$ and by admitting a
conformal
motion parallel to the 4-velocity. We provide explicit forms of the state
variables and equations of state linking them. These formal thermodynamic
relations are determined up to an arbitrary function of time which reduces
to the FRW scale factor in the FRW limit of the solutions. We show that a
formal identification of this free parameter with a
FRW scale factor determined by FRW dynamics leads to an unphysical temperature
evolution law. If this parameter is not identified with a FRW scale factor,
it is possible to find examples of solutions and formal equations of state
complying with suitable energy conditions and reasonable asymptotic behavior
and temperature laws. \Par
\vskip1cm\ni
\vskip 4cm
\ni $^*$Work supported by CONACYT, M\'exico, project No. 3567-E

\vfill
\eject

\baselineskip 18pt

\ni {\bf I. Introduction} \par
\vskip .3cm

The \qleft Stephani Universes\qright is the generic name for a class of
metrics comprising the most general conformally flat solution with an
irrotational perfect fluid source$^{1-6}$. These solutions, admitting
(in general) no isometries, generalize FRW spacetimes (their particular case
 with vanishing
4-acceleration), and could have a physical interest as simple inhomogeneous
and anisotropic cosmological models. However, it is well known that Stephani
Universes (except for the FRW subcase) do not admit a
barotropic equation of state $p=p(\rho)$, where $p$ and $\rho$ are the pressure
and matter-energy density, perhaps explaining why so
few references are found in the literature$^{3,7,8}$ studying the
physics of (non-FRW) Stephani Universes. However,  barotropic
equations of state might be too restrictive$^9$, and so we aim in
this paper to verify whether one can find more arguments to gauge the
physical viability of these solutions besides simply dismissing them for
not admitting a barotropic equation of state.

Coll and Ferrando$^{10}$ addressed the question
of the consistency of the thermodynamical equations with Einstein field
equations, deriving rigurously the criterion to verify if a single component
perfect fluid source of a given exact solution admits what these authors
denote a \qleft thermodynamic scheme\qright.
However, Coll and Ferrando did not go
beyond the admisibility of their consistency criterion, that is, into the
physics of the fluids: note that it is perfectly possible to have
unphysical fluids whose thermodynamics is formally
correct. In a recent paper$^{11}$, we have expanded and
complemented the work of Coll and Ferrando by applying
their criterion to irrotational, non-isentropic and
geodesic fluids (the perfect fluid Szekeres solutions). Regarding the Stephani
Universes, Bona and Coll$^7$
did apply the work of Coll and Ferrando to these solutions, claiming that
non-barotropic cases only admit a thermodynamic scheme if $r\ge 2$ and
presenting a very brief discussion of the
thermodynamics of non-barotropic Stephani Universes with $r= 2$.
By providing a specific counterexample, we prove in this paper that the
result of Bona and Coll is incorrect (see section V).
We also provide a deeper discussion of the thermodynamics of
non-barotropic Stephani Universes with $r<2$ admitting a thermodynamic scheme.
It is important to specify that the study of this thermodynamics assumes
the matter source to be a single component perfect fluid. If the source
were a mixture of perfect fluids, the study of its thermodynamics would
involve looking at chemical potentials, and so would be an altogether
different problem, a problem which will not be addressed in this paper.
We assume henceforth that all mention of the term fluid indicates a
single component fluid. The contents of this paper are described below.

We present in section II a summary of the equations of the
thermodynamics of a general relativistic
perfect fluid, together with the conditions for admissibility of a
thermodynamic scheme. We re-phrase the conditions
derived by Coll and Ferrando in terms of differential forms expanded
in a coordinate basis adapted to the comoving frame in which the Stephani
Universes are usualy described. These conditions
are applied in section III to the Stephani
Universes, yielding an interesting result: these solutions do not admit
a thermodynamic
scheme in general, that is, with unrestricted values of their free parameters.
However, under
suitable restrictions of these parameters, we find a specific
subclass of non-barotropic Stephani Universes which does
admit a thermodynamic scheme when $r< 2$. This subclass is the counter
example to the work of Bona and Coll$^7$ mentioned above.

In section
IV, we derive for the non-barotropic Stephani Univeres complying with a
thermodynamic scheme
explicit expressions of all state variables: $\rho$ and $p$,
particle number density $n$, specific entropy $S$ and temperature $T$, as well
as two-parameter equations of state linking them. The latter turn out to be
difficult to interpret as there is no clue on how to fix the only time
dependent free parameter of the solutions and its relation with the matter
energy density. We explore in section V the strategy which consists
in formally identifying these quantities with the scale factor and
matter-energy
density of a FRW limiting spacetime assumed to comply with a \qleft gamma
law\qright equation of state. The resulting temperature evolution law is
unphysical and does not reduce in the FRW limit to that expected for a
FRW cosmology with such an equation of state. On the other hand, if this
identification with FRW parameters is abandoned, we show in section VI that
it is possible to
obtain temperature evolution laws and equations of state which, being
more formal than physical, are not altogether unphysical and do comply with
suitable energy conditions. Conclusions are
presented and summarized in section VII.

We prove in the Appendix that Stephani Universes admitting a thermodynamic
scheme are characterized invariantly by admitting: (a) a spacelike Killing
vector ($r=1$) and (b) a conformal Killing vector field parallel to the
4-velocity. In fact, we
show that these Stephani Universes comprise
the most general class of solutions with an irrotational perfect fluid
source admitting this type of conformal symmetry.

\Par

\vskip .3cm
\ni {\bf II. Thermodynamics of a non-isentropic irrotational perfect fluid.}
\par
\vskip .3cm

Consider the energy--momentum tensor for a perfect fluid

$$T^{ab}=(\rho+p)u^au^b+ p g^{ab}\eqno(1)\qquad$$

\ni where $\rho$, $p$ and $u^a$ are the matter-energy density, pressure and
4-velocity, respectively. This tensor satisfies the conservation law
$T^{ab}_{\ ;b}=0$ which implies the contracted Bianchi identities

$$\dot\rho+(\rho+p)\Theta=0\eqno(2a)\qquad$$

$$h_a^b p,_b+(\rho+p)\dot u_a=0\eqno(2b)\qquad$$

\ni where $\Theta=u^a_{;a}$, $\dot u_a=u_{a;b}u^b$
and $h_a^b=\delta_a^b+u_au^b$ are respectively the expansion, 4-acceleration
and projection tensor and $\dot\rho=u^a\rho,_a$. The thermodynamics of a
perfect fluid is
essencially contained in the matter conservation law, the condition of
vanishing entropy production and the Gibbs-Duhem relation. The first two are
given by

$$(nu^a)_{;a}=0\eqno(3a)\qquad$$

$$(nSu^a)_{;a}= 0\eqno(3b)\qquad$$

\ni where $n$ is the particle number density and $S$ is the specific entropy.
Condition (3a) inserted in (3b) leads to $u^aS,_a=\dot S=0$, so that $S$ is
conserved along the fluid lines but is not a universal constant. In the latter
case we have: $\d S=0$, and the fluid is isentropic, admitting a barotropic
equation of state. The Gibbs-Duhem
relation can be given as the 1-form

$$\w=\d S={1\over{T}}\left[ \d \left({{\rho}\over{n}}\right)+p\d
\left({1\over{n}}\right)\right]\eqno(4)\qquad$$

\ni where $T$ is the
temperature. The necessary and sufficient condition for the integrability
of (4)

$$\w\wedge \d\w=0\qquad\hbox{necessary and sufficient}\eqno(5)\qquad$$

\ni subjected to
fulfilment of the conservation laws (2) and (3), are the conditions which
Coll and Ferrando denote admissibility of a \qleft thermodynamic scheme\qright.
These conditions were given by these authors as

$$(\dot p \d\dot\rho -\dot\rho\d\dot p)\wedge\d p\wedge\d \rho
=0\eqno(6)\qquad$$

\ni Another integrability condition, not examined by Coll and Ferrando, is

$$\d\w=0\qquad\hbox{sufficient}\eqno(7)\qquad$$

The perfect fluid source of the Stephani Universes is characterized by an
irrotational (hence, hypersurface orthogonal) and shear-free 4-velocity.
For such a fluid source there exist$^{2,4,5}$
local comoving coordinates $(t, x^i)$, such that the metric, 4-velocity,
4-acceleration, expansion and projection tensor are given by

$$ds^2=-N^2 dt^2+L^2\delta_{ij}dx^i dx^j\eqno(8a)\qquad$$

$$N={{L_{,t}/ L} \over {\Theta / 3}}$$

$$u^a=N^{-1}\delta^a_t\qquad \dot u_a=(\log N)_{,a}\delta^a_i
\eqno(8b)\qquad$$

$$h_{ab}=g_{ij}\delta_a^i\delta_b^j\eqno(8c)\qquad$$

\ni where $\Theta=\Theta(t)$ and the metric function $L$ is (in general)
a function of all the coordinates $(t,x^i)$. In this representation $\dot X
=(1/N)X_{,t}$ for all scalar functions and the
Bianchi identities and conservation laws (2) and (3) become

$$\rho_{,t}+(\rho +p)(\log n)_{,t} =0\eqno(9a)\qquad$$

$$p_{,i}+(\rho +p)(\log N)_{,i} =0\eqno(9b)\qquad$$

$$n={{n_0(x^i)} \over {L^3 }}\eqno(9c)\qquad$$

$$S=S(x^i)\eqno(9d)\qquad$$

\ni where $n_0(x^i)$ appearing in (9c) is an arbitrary
function denoting the conserved particle number distribution. In the
the coordinate basis
of 1-forms $(\d t,\d x^i)$ associated with the comoving frame (8),
the Gibbs-Duhem relation reads

$$\w =S_{,i}\d x^i={1 \over T}\left[ {\left( {{\rho  \over n}}
 \right)_{,i}+p\left( {{1 \over n}} \right)_{,i}} \right]\d
x^i\eqno(10)\qquad$$

\ni where the $t$ component of $\w$ in this coordinate basis vanishes
due to (9d). A sufficient integrability condition of (10) is given by

$$\d\w=W_{ti}{{\d t\wedge \d x^i} \over {nT}}+W_{ij}{{\d x^i\wedge \d x^j}
 \over {nT}}=0\eqno(11)\qquad$$

$$W_{ti}={{p_{[,i}n_{,t]}-n^2T_{[,t}S_{,i]}} \over n}=
\left( {\rho +p} \right)\left( {{{n_{,i}} \over n}{{T_{,t}} \over T}
-\dot u_i{{n_{,t}} \over n}} \right)-\left( {\rho _{,i}{{T_{,t}}
\over T}+p_{,t}{{n_{,i}} \over n}} \right)$$

$$W_{ij}={{p_{[,i}n_{,j]}+n^2T_{[,i}S_{,j]}} \over n}=
{{T_{[,i}\rho _{,j]}} \over T}-\left( {\rho +p} \right)
\left( {{{T_{[,i}+T\dot u_{[i}} \over T}} \right){{n_{,j]}} \over n}$$

\ni where square brackets denote antisymmetrization on the corresponding
indices. The necessary and sufficient condition (5) is given by

$$\d\w\wedge \w=X_{ijk}{{\d x^i\wedge \d x^j\wedge \d x^k}
\over {n^3T^2}}+X_{tij}{{\d t\wedge \d x^i\wedge \d x^j} \over {n^3T^2}}=0
\eqno(12)\qquad
$$

$$X_{ijk}=-\rho _{[,i}p_{,j}n_{,k]}$$

$$X_{tij}=\rho _{[,t}p_{,i}n_{,j]}$$

\ni Conditions (12) are entirely equivalent to (6) provided by
Coll and Ferrando. One can obtain the latter form the former simply by using
(2) and (3) (in their forms (9)). However, (11) and (12) are more intuitive
than (5) and (6), as they directly incorporate state
variables such as $n$, $S$ and $T$, and their relations with $\rho$ and $p$.
Condition (12) is also more practical than (6), as it is easier to use $n$
and $S$ from (9c) and (9d) than to compute the set $(\rho,\dot\rho,p,\dot p)$
in exact solutions in which these quantities can be quite cumbersome.
The sufficient condition (11), not examined by Coll and Ferrando, is also
helpful, since if its fulfilment guarantees that (6) (or (12)) holds.

As shown in the following section, if a solution of Einstein equations
is available (thus providing
$\rho$ and $p$ in terms of the metric functions) it is straightforward to
verify the admissibility of the thermodynamic scheme. This we will
do for the Stephani Universes which are particular cases of (8), and to do
so we suggest
the following procedure: (a) solve the conditions (12) and substitute the
solution into (10), thus identifying possible (non-unique) forms for $S$ and
$T$; insert the obtained forms of $T$ and $n$ into (11) in order to verify if
further restrictions follow from the sufficient conditions.
If these conditions
hold, the equations of state linking the state variables $(\rho,p,n,S,T)$
(together with their functional relation with respect to the
metric functions) follow directly from integrating them.

\par

\vskip .3cm
\ni {\bf III. The Stephani Universes.} \par
\vskip .3cm

The Stephani Universes are described by the particular case of (8a)
given by:

$$L(t,x^i)={{R} \over {1+2A_ix^i+(A^2+(k/4)R^2)\delta_{ij}x^ix^j}}
\eqno(13)\qquad$$

\ni  with $A_i(t)= (A_x(t),
A_y(t), A_z(t))$, $A^2\equiv \delta^{ij}A_iA_j$, $k(t)$ and $R(t)$ are
arbitrary
functions. Notice that the Stephani Universes contain FRW spacetimes as the
particular case $A_i=0$, $k=k_0=\hbox{const.}$ in (8) and (13). As it is well
known that
the latter are the subclass of barotropic Stephani Universes, we assume
hereafter (and unless stated otherwise) that all mention of Stephani Universes
excludes their FRW subclass. The field equations associated with (8) and (13)
are

$$\rho =\rho (t)={{\Theta ^2} \over 3}+3k\eqno(14a)\qquad$$

$$p=-\rho -{{\rho ,_t} \over{3L,_t/L}}\eqno(14b)\qquad$$

\ni where the contracted Bianchi identity (9a) has been used as a definition
of $p$. The remaining Bianchi
identities and conservation laws are given by
equations (9b-d), while the Gibbs-Duhem  1-form (10) and the
integrability conditions (11) and (12) follow as

$$S,_i={{(\rho +p)}\over{T}} \left( {{1 \over n}}
\right)_{,i}\eqno(15)\qquad$$

$$W_{ti}=\left[ {(\rho +p){{T,_t} \over T}-p,_t}\right]n,_i-(\rho +p)\dot
u_in,_t=0\eqno(16a)\qquad$$

$$W_{ij}=0\Rightarrow \Omega_{[i}(\log n)_{,j]}=0\eqno(16b)\qquad$$

\ni where

$$\Omega _i\equiv (\log T)_{,i}+\dot u_i=(\log TN)_{,i}
\eqno(16c)\qquad$$

$$X_{tij}=0\Rightarrow p_{[,i}n_{,j]}=0\Rightarrow
(\log N)_{[,i}(\log n)_{,j]}=0\eqno(17)\qquad$$

\ni Inserting the forms of $N$ and $n$ given by (8a) (9c) and (13) into the
necessary and sufficient condition (17) yields a general solution of the
latter given by

$$a(t)\log\left( {{L_{,t}}\over{L}}\right)+b(t)\log\left({{f}\over{L^3}}
\right)=\log (c(t))\eqno(18)\qquad$$

\ni where $(a,b,c)$ are arbitrary functions. This integrability condition is
not satisfied in general, that is for arbitrary forms of the free functions
$(A_i, k, R)$ appearing in the metric function (13). Bona and Coll$^7$ have
claimed that this condition is only
satisfied by Stephani Universes with isometry groups of dimension $r\ge 2$.
However, it is possible to provide a particular case of this
metric (characterized by $r=1$, see Appendix) which satisfies (16) and
(17) and so, leads to well defined forms for $T$ and $S$. This case is
characterized by the existence
of a conformal Killing vector field parallel to the 4-velocity
(see Appendix), the corresponding forms of $A_i$ and $k$ are

$$A_i=a_iR,\qquad k={k_0\over{R^2}}+{b_0\over{R}}-4\delta^{ij}a_ia_j
\eqno(19a)\qquad$$

\ni where $(a_i, k_0, b_0)$ are arbitrary constants and $R$ remains
arbitrary. With these parameter values, (18) holds with

$$a(t)=1,\quad b(t)=1/3,\quad c(t)=-(1/R)_{,t}\eqno(19b)\qquad$$

$$n_0=f^{-3}\quad \hbox{where:}\quad f\equiv 1+{\textstyle{{1} \over 4}}
k_0\delta _{ij}x^ix^j\eqno(19c)\qquad$$

\ni Inserting (19a) and (19c) into (15) yields

$$TS_{,i}=-3(\rho +p)(Lf)^4(F/ f)_{,i}\eqno(20a)\qquad$$

\ni where

$$F =2a_ix^i+{\textstyle{1 \over 4}}b_0\delta _{ij}x^ix^j
\eqno(20b)\qquad$$

\ni This equation shows how non unique forms of $S$ and $T$ emerge
 if we demand only the fulfilment of condition (17) (or (18)). A
general expression for these quantities, in agreement with (15), is given by

$$S=S(\sigma)\qquad \sigma\equiv -{F\over{f}}\eqno(21a)\qquad$$

$$T={{3(\rho +p)} \over {S'n^{4/ 3}}}\eqno(21b)\qquad$$

\ni where a prime denotes derivative with respect to
$\sigma$ and (9c) (13) have been
 used to eliminate $L$ in terms of $n$.
Regarding the sufficient conditions (16), $T$ given by (21b) satisfies (16a).
This is easily verified by
eliminating $\dot u_i$ from (16c) and inserting this result together with
$T_{,t}/T$  computed from (21b) into (16a). On the other hand, inserting
(21b) into (16b) and using (17) leads to the condition
$S''=0$, so that $S$ is a linear function of $\sigma$. This yields
the following forms for $T$ and $S$

$$S=S_0+\sigma\eqno(23a)\qquad$$

$$T={{3(\rho +p)} \over{n^{4/ 3}}}\eqno(23b)\qquad$$

\ni which are compatible with both sets of conditions (16) and (17).A
discussion on the interpretation of the expressions derived above is provided
in the following sections.

\Par

\vskip .3cm
\ni {\bf IV. Formal equations of state.} \par
\vskip .3cm

The particular subclass of Stephani Universes complying with the
thermodynamic scheme, as characterized by the parameter restrictions (19a),
is described by the conformally FRW metric

$$ds^2=\Phi ^2\left[ {-dt^2+{{R^2\delta _{ij}dx^idx^j} \over
 {\left( {1+{\textstyle{{1} \over 4}}k_0\delta _{ij}x^ix^j}
 \right)^2}}} \right]\eqno(24a)\qquad$$

$$\Phi ={{1+{\textstyle{{1} \over 4}}k_0\delta _{ij}x^ix^j}
\over { {1+{\textstyle{{1} \over 4}}k_0\delta _{ij}x^ix^j
+R\left( {2a_ix^i+{\textstyle{{1} \over 4}}b_0\delta _{ij}x^ix^j}
 \right)} }}={f \over {f+RF}}\eqno(24b)\qquad$$

\ni Its corresponding field equations are

$${\textstyle{1 \over 3}}\rho =\left( {{{R_{,t}} \over R}}
 \right)^2+{{k_0} \over {R^2}}+{{b_0} \over R}-4\delta ^{ij}a_ia_j
\eqno(25a)\qquad$$

$$p=-\rho -{\textstyle{1 \over 3}}R\rho _{,R}(1-R\sigma )
\eqno(25b)\qquad$$

\ni where we have chosen $\Theta/3=R_t/R$ and $\sigma$
is given by (21a).  These Stephani Universes
can be characterized invariantly by the existence of a
conformal Killing vector field parallel to the 4-velocity (see Appendix).
Assuming the forms
of $T$ and $S$ from (23), the term $\sigma$ appearing in (25b) (and
in other expressions) is a linear function of $S$. Also, since
$\rho$ and $R$ are both functions of $t$, all terms involving $R$ and
$\rho_{,R}$ can be expressed as functions of $\rho$. This results in
the following forms for generic equations of state, expressing $p$, $n$ and
$T$ in terms of $\rho$ and $S$

$$p(\rho ,S)=-\rho -{\textstyle{1 \over 3}}R\rho _{,R}\left[
{1+(S-S_0)R} \right]\eqno(26a)\qquad$$

$$n(\rho ,S)=\left[ {{1 \over R}+S-S_0}
 \right]^3\eqno(26b)\qquad$$

$$T(\rho ,S)={{-3R^5\rho _{,R}} \over {\left[ {1+(S-S_0)R}
 \right]^3}} \eqno(26c)\qquad$$

\ni These are {\it {formal}} equations of state, in the sense that they
are not dictated by physical considerations and imposed before solving the
constraints of the field equations ({\it {physical}} equations of state), but
arise as a consequence of imposing the fulfilment of the thermodynamic scheme
on metric functions whose spacial dependence has been fixed by imposing
conformal flatness (a geometric constraint: vanishing of the Weyl tensor).
The best one can do in this case is to verify if these formal thermodynamic
relations could be manipulated in such a way that solutions (24) could
describe physically reasonable cosmologies. However, equations (25) are still
undetermined: a choice of $\rho(R)$ must be made in order to determine these
equations and to be able to integrate the Friedmann-like equation (25a)
to yield the time evolution of the metric. Unfortunately, there is no clear
cut way guiding one on how to select $\rho(R)$. Various possibilities are
explored below.

\Par

\vskip .3cm
\ni {\bf V. FRW limit.} \par
\vskip .3cm

The metric (24) bears a close resemblance to a FRW metric. As equations
(24)-(25) reveal, the term
$\sigma$, related to the spacially
dependent entropy density, is the term that makes these solutions
inhomogeneous and anisotropic ({\it {i.e.}} non-FRW). In fact,
their FRW limit follows if $b_0\to 0$ and $a_i\to 0$, so that
$\sigma\to 0$. Under this limit, the metric and field equations (25)
become the metric and field equations of a perfect fluid FRW spacetime
with scale factor
$R$ and $k_0=0,\pm1$ marking the curvature of the spacial
sections. This correspondence and resemblance to FRW
cosmologies motivates us
to verify if these solutions could be considered as some sort of
\qleft near-FRW\qright cosmologies and if they could be examined
within the framework of
a FRW limit. Hence we pose the question of
whether the state variables and their equations of state become
those one would expect of a \qleft near-FRW\qright cosmology if
we assume that $R$ can be fixed as it were a FRW
scale factor. Along these lines, we notice that the state variables
$\rho$, $p$ and $n$, given by (25a), (25b), (26a) and (26b), also tend to
their FRW values, but the limiting form of $T$ in (26c) takes the strange form
$T\to -R^5\rho_{,R}$. In order to see what sort of temperature law and
equations of state correspond
to this FRW limit, we assume a \qleft gamma law\qright
equation of state $p_0=(\gamma -1)\rho$, where $p_0$ is given by setting
$S=S_0$ in (26a), the form of $\rho(R)$ becomes

$$\rho (R)=\left( {{R \over {R_0}}} \right)^{-3 \gamma}$$

\ni and so the various forms (26) of the equation of state become

$$p(\rho ,S)=(\gamma -1)\rho +{\textstyle{1 \over 3}}\gamma R_0\rho ^{1-1/
3\gamma }(S-S_0)\eqno(27a)\qquad$$

$$n(\rho ,S)=\left[ {\rho ^{1/ 3\gamma }+S-S_0}
 \right]^3\eqno(27b)\qquad$$

$$T(\rho ,S)={{\gamma R_0^4 \rho ^{1-1/ 3\gamma }} \over
{\left[ {\rho ^{1/ 3\gamma }+S-S_0}
 \right]^3}}\eqno(27c)\qquad$$

\ni Irrespective of the interpretation of these strange formal thermodynamic
relations, notice that $p$ and $n$ do reduce to their FRW values in the FRW
limit, though $T\to \rho^{1-4/3\gamma}\propto R^{4-3\gamma}$, a temperature
evolution law which has no relation to that expected at the FRW limit: for
$\gamma=1$ (dust), $T\propto R$ instead of $T=\hbox{const.}$ and for
$\gamma=4/3$ (radiation), $T\propto \hbox{const.}$ instead of $T\propto
R^{-1}$. Therefore, this formal identification of $R$ with the FRW scale
factor leads to contradictory results.
Of course, one could also consider $R$ as a FRW scale factor associated
with other equations of state, however, the presence of the strange term
$-R^5\rho_{,R}$ in (26c) makes it highly improbable for this temperature
evolution law to have any meaningful correspondence with that of a limiting
FRW cosmology.

\par

\vskip .3cm
\ni {\bf VI. Examples complying with minimal physical requirements.} \par
\vskip .3cm

Looking at the equations of state (26) as formal
thermodynamic relations, which hopefuly might provide at least
a gross approximation to physical relations, and if one is prepared to abandon
the identification of $R$ with a specific FRW scale
factor, it is possible at least to verify if
the free parameters $(a_i, k_0, b_0)$ and $R(t)$ can be selected to
assure that solutions (24) and equations (26) comply with minimal physical
conditions, such as energy conditions and an acceptable asymptotic behavior.
In order to examine equations (24)-(26) within this framework, consider the
coordinate transformation

$$\eqalign{&x=2\Gamma (\chi / 2)\sin \theta \cos \varphi \cr
  &y=2\Gamma (\chi / 2)\sin \theta \sin \varphi \quad\quad\Gamma(\chi/2) =
\left\{ \matrix{\tan(\chi/2) \quad k_0=1\hfill\cr
  \tanh(\chi/2) \quad k_0=-1\hfill\cr} \right.\cr
  &z=2\Gamma (\chi / 2)\cos \theta \cr}$$

\ni which brings the metric (24) into the simple form

$$ds^2={{-dt^2+R^2\left[ {d\chi ^2+\Sigma ^2(\chi )(d\theta ^2+
\sin ^2\theta d\varphi ^2)} \right]} \over {\left[ {1+R\left(
{2\Sigma (\chi )V(\theta ,\varphi )+b_0\Sigma ^2(\chi / 2)} \right)}
 \right]^2}}\eqno(28a)\qquad$$

$$\Sigma(\chi) =\left\{ \matrix{\sin\chi \quad\quad k_0=1\hfill\cr
  \sinh \chi \quad\quad k_0=-1\hfill\cr} \right.\eqno(28b)\qquad$$

$$V(\theta ,\varphi )=a_1\sin \theta \cos \varphi +a_2\sin
\theta \sin \varphi +a_3\cos \theta \eqno(28c)\qquad$$

The state variables in (26) are unaffected by this coordinate transformation,
with $\sigma$ and the function $n_0$ in
(19c) which provides the conserved particle number at an inicial hypersurface
$t=\hbox{const.}$ now given by

$$\sigma=2\Sigma (\chi )V(\theta ,\varphi )+b_0\Sigma ^2(\chi / 2)
\eqno(29a)\qquad$$

$$n_0=\left\{ \matrix{\cos ^6({\textstyle{1 \over 2}}\chi )\quad k_0=1\hfill\cr
  \cosh ^6({\textstyle{1 \over 2}}\chi )\quad k_0=-1\hfill\cr} \right.
\eqno(29b)\qquad$$

\ni From (28c), the term $V$ containing the dependence on $(\theta,\varphi)$
is bounded, hence $\sigma=-F/f$ might only diverge along $\chi\to\pm\infty$
(irrespective of the value of $\rho=\rho(t)$) for the case $k_0=-1$.
This fact can be examined from another angle: the function $\sigma$
(and so, the entropy density $S$) can be expressed in terms of $n_0$
as

$$S=S_0+\sigma=S_0+4n_0^{1/6}(1-k_0 n_0^{1/3})^{1/2}+b_0(1-k_0
 n_0^{1/3})\eqno(29c)\qquad$$

\ni and so, for $k_0=-1$, the initial particle number $n_0$  becomes infinite
as $\chi\to\pm\infty$, this infinite concentration of particles  causes
the entropy density $S$ and the pressure to diverge along these limits.
This means that solutions with $k_0=-1$ have an undesirable asymptotic
behavior. On the other hand, the locus $R=0$
marks another singularity, analogous to a FRW big bang, while the vanishing
of the denominator in (28a) indicates an asymptotically deSitter evolution
($p\to -\rho$, $n\to\infty$) characterized by the unphysical behavior $T\to
\infty$. Therefore, for the formal equations of state (26)
to have any physical meaning, we must choose
$k_0=1$ and demand the condition $1-R\sigma\ne 0$ to hold, together
with the dominant and weak energy conditions which can be
combined into the restriction: $0\le p/\rho\le 1$.
Also, all state variables must diverge at the big-bang singularity $R=0$.
 From equations (26), the conditions $0\le p/\rho\le 1$ and  $T\to\infty$ hold
at the limit $R\to 0$ if $\rho\approx R^{-(4+m)}$ for $m>0$ at this limit,
leading to the following asymptotic values:

$${p \over \rho }\approx {\textstyle{1 \over 3}}(m+1)\eqno(30a)\qquad$$

$$n\approx {1 \over {R^3}}\approx \rho ^{3/ (m+4)}\eqno(30b)\qquad$$

$$T\approx {{m+4} \over {R^m}}\approx (m+4)\rho ^{m/ (m+4)}\eqno(30c)\qquad$$

\ni  as $R\to 0$. Since $R$ is no longer constrained to be interpreted
as a sort of FRW scale factor in a FRW limit, we can devise
a simple example complying with the
conditions mentioned above and the asymptotic limits (30)
by choosing a simple power law $\rho=(R_0/R)^6$, where $R_0$ is a constant.
This choice leads to the following state variables

$$p(\rho ,S)=\rho +2R_0(S-S_0)\rho ^{5/ 6}\eqno(31a)\qquad$$

$$n(\rho ,S)=\left[ {\rho ^{1/ 6}+S-S_0} \right]^3
\eqno(31b)\qquad$$

$$T(\rho ,S)={{6R_0^4 \rho ^{5/ 6}} \over {\left[ {\rho ^{1/ 6}+
S-S_0} \right]^3}}\eqno(31c)\qquad$$

\ni which yield a stiff fluid equation of
state  $p/\rho\to 1$ and $T\to\infty$ near the big-bang singularity ($R=0$
and/or $\rho\to\infty$). However, as $R\to\infty$ (or $\rho\to 0$), the
dominant
energy condition could be violated: $p/\rho\to\infty$.  This can be avoided by
selecting the remaining
arbitrary constants $(a_i, b_0)$ in such a way that $S _0 >S$ and
$1-R\sigma=1+ R(S-S_0)> 0$ holds everywhere and the Friedmann-like
equation

$$\left( {{{R_{,t}} \over R}} \right)^2={{R_0^6} \over {3R^6}}-{1 \over {R^2}}
-{{b_0} \over R}+4\delta ^{ij}a_ia_j\eqno(32a)\qquad$$

\ni obtained by substituting $k_0=1$ and $\rho=(R_0/R)^6$ into (25a), has no
solutions $R(t)$  allowing for $R\to\infty$, or equivalently,
that $\rho$ does not vanish along the time evolution of the fluid. These
conditions require (32a) to have a real positive root and

$${{R_{,tt}} \over R}=-{2 \over 3}\left( {{{R_0} \over R}}
 \right)^6-{{b_0} \over {2R}}+4\delta ^{ij}a_ia_j<0\eqno(32b)\qquad$$

\ni along this root, hence $R(t)$ is convex. If we assume $R=R_0>0$ to be the
value along which $R_{,t}/R=\Theta/3$ vanishes, equation (32a) fixes $R_0$ in
terms of the parameters $(a_i, b_0)$ as the positive root of:

$$({\textstyle{1 \over 3}}+4\delta ^{ij}a_ia_j)R_0^2-b_0R_0-1=0
\eqno(32c)\qquad$$

\ni which substituted into (32b) yields the condition of convexity.
It is not difficult to find combinations of parameters $(a_i, b_0)$
so that the fluid has the desired type of kinematic evolution (qualitatively
analogous to a standard \qleft closed\qright FRW cosmology) and
physically correct behavior of the state variables: that is,
to have the ratio $p/\rho=1$ and $T\to\infty$ at
the big bang evolving to values in the range $0\le p/\rho<1$ and cooling to
$T$ finite as the fluid reaches its maximum expansion, bounces and then
recollapses with $p/\rho=1$ and $T\to\infty$ at the big crunch.
Such an example is illustrated in figure 1.

\Par

\vskip .3cm
\ni {\bf VII. Conclusions.} \par
\vskip .3cm

We have investigated the consistency of the thermodynamic equations
following from the condition of existence of a thermodynamic scheme
for the Stephani Universes,
whose source is a non-isentropic perfect fluid (thus, not admitting a
barotropic equation of state). This work has aimed
at improving the study of this type of
solutions, as classical fluid models generalizing FRW cosmologies,
in contrast to a widespread attitude of simply disregarding
them for not admitting a barotropic equation of state.

For the particular subclass of Stephani Universes admitting a thermodynamic
scheme, the resulting equations of state have an ellusive interpretation, as
there is no blue print on how to select the time dependent free parameter of
the solutions, an arbitrary function $R$ reducing to the FRW scale factor in
the FRW limit.
We have shown that by formally identifying this parameter with the FRW scale
factor of a FRW cosmology satisfying a \qleft gamma law\qright equation
of state leads to an unphysical temperature evolution
law, totally unrelated to that of their limiting FRW cosmology. The question
of how to select these parameters in a convenient way remains unsolved,
though the adequate theoretical framework to carry this task has been
presented in section VI. We have shown that combinations of free parameters
exist so that the formal equations of state comply with minimal physical
requirements.

We have shown that Stephani Universes (other
than the FRW subclass or the subclass presented in previous sections)
are not compatible, in general, with a thermodynamic scheme. This
fact seems to disqualify these solutions as classical fluids of physical
interest. However, the latter can still be useful if they are
examined under a less restrictive framework than that of the simple
perfect fluid. In this context, the Stephani Universes (like other
perfect fluid solutions with a shear-free 4-velocity) can be recast as
exact solutions for a fluid with a bulk viscous
stress$^{12}$. From this point of view, the thermodynamics is not only
totally different as that of the perfect fluid case but
much less restrictive and more amenable to satisfy the criteria for
constructing inhomogeneous and anisotropic cosmological models of physical
interest.

\Par

\vskip .3cm
\ni {\bf Acknowledgements} \par
\vskip .3cm
We are grateful to Kayll Lake for providing us with a copy of the tensor
package GRTensor$^{13}$ running with the algebraic computing
language Maple V$^{14}$. All calculations in the paper were
verified with this package and with the help of Maple V symbolic
computing functions.
We also wish to thank Luis N\'u\~nez and Ricardo Bartolotti
from Universidad de los Andes, M\'erida, Venezuela, for their hospitality
 and for verifying the components of the Killing
equation for the metric (24) with a program they wrote in Maple V. One
 of us (R.A.S.) acknowledges H.\ Koshko, N.\ Pochol
and F.D.\ Chichu for critically
meowing their comments on the manuscript. \Par

\Par

\vskip.5cm
\ni {\bf Appendix. Isometries and conformal symmetries.}
\vskip .3cm

\ni  {\bf{ (a) The metric (24) admits a one-parameter group of isometries.}}
\vskip .3cm

Being conformally
flat, (24) admits a $G_{15}$ of conformal symmetries. Various isometry
subgroups are readily identified. If $k_0=0$, irrespective of the
values of the remaining constants $a_i$, (24) becomes
spherically symmetric (this can be verified through re-scalings of
the form $x^i=\bar x^i+c^i$). Consider now (24) with $k_0=\pm1$ and $a_i$
arbitrary nonzero constants. Since we are assuming this spacetime to be
non-static ($\Theta=3R_{,t}/R\ne 0$), its Killing
vectors (if they exist) must all be tangent to the hypersurfaces orthogonal
to the 4-velocity$^{15}$. Spherical symmetry is easily identified by setting
$a_i=0$ and $b_0\ne 0$, while FRW subcases follow from $a_i=0=b_0$.
However, for general values of these constant parameters,
it is not easy to identify at first glance the existence of isometries groups
$G_1$ or $G_3$ acting along the hypersurface orthogonal to the 4-velocity.

In order to find out if (24) admits isometries, and if so, the dimension of
their orbits, we follow the result of theorem 3 in Bona and Coll$^6$.
Identifying carefully the parameters used by these authors: $a$, $b$
and $\phi$ (see their equations (2)-(4)) with the corresponding
parameters in (8) and (13), we find that their function $a$ is our function
$R$, their vector $b$ is
the vector formed by the functions $A_i$ in (13), while $\phi=A^2+(k/4)*R^2$,
that is, the coefficient of the quadratic term in the denominator of (13).
The case complying with the thermodynamic scheme (the metric (24)), is
defined by (19a), thus we have for this metric $b=(a_1, a_2, a_3)R$ and
$\phi=b_0R$. Therefore, metric (24) with arbitrary values of its free
parameters $R$, $a_i$, $b_0$, corresponds to the case: $\dot b\ne 0$,
$\ddot b\ne0$ with $\dot b\wedge \ddot b=0$ and with
$\dot\phi=b_0\dot R\ne 0$. According to the classification of
page 616 of Bona and Coll$^6$, the metrics (24) admit a one parameter
isometry group with $r=1$, a case these authors identify as axially
symmetric.

\par
\vskip .3cm
\ni {\bf { (b) All Stephani Universes associated with the metric (24)
 admit a conformal Killing vector parallel to
the 4-velocity.}}

\vskip .3cm

We have proven that the metric (24) describes a class of
Stephani Universes admitting a thermodynamic scheme and having an isometry
group of dimension $r< 2$. On the other hand,
it is well known$^{16}$ that the 4-velocity in FRW spacetimes
($u_{_{(0)}}^a=\delta^a_t$ in the comoving coordinates of (24))
is a conformal Killing vector, satisfying $\xi_{(a;b)}=\psi_{_{(0)}} g^{^{(0)}}
_{ab}$ with scale factor $\psi_{_{(0)}}=R_{,t}$.
Since (24) is conformally related to a FRW metric, the vector field
$\xi^a=\delta^a_t$ in (24) (parallel to the 4-velocity $u^a=(-\Phi)^{-1}
\delta^a_t$) is a conformal Killing vector $\xi^a$ in (24), satisfying
$\xi_{(a;b)}=\psi g_{ab}$, with conformal factor $\psi$ given by$^{16}$

$$\psi=\psi_{_{(0)}}+\xi^a(\log \Phi)_{,a}=R_{,t}\left(1-{F\over{f+RF}}\right)
\eqno(A1)\qquad$$

\Par
\vskip .3cm
\ni  {\bf{ (c) The metric (24) describes the most general perfect fluid
spacetime admitting a conformal Killing vector parallel to the 4-velocity.}}
\vskip .3cm

It is known$^{17}$ that
the existence of a conformal symmetry of this type ( $\xi^a=\Omega u^a$,
satisfying $\xi_{(a;b)}=\psi g_{ab}$) requires the fluid
4-velocity to be shear-free, with the remaining kinematic parameters
given by

$$\dot u_a=u_{a;b}u^b=h_a^b(\log\Omega)_{,b}\eqno(A2)\qquad$$

$$\omega _{ab}=u_{[a;b]}+\dot u_{[a}u_{b]}=\Omega (\Omega ^{-1}\xi
 _{[a})_{;b]}\eqno(A3)\qquad$$

$${\textstyle{1 \over 3}}\Theta ={\textstyle{1 \over 3}}u^a_{;a}
=u^a(\log \Omega )_{,a}={\psi  \over \Omega }\eqno(A4)\qquad$$

\ni Consider the irrotational case $\omega _{ab}=0$, as described by the
comoving coordinates of (8) but with $L$ not necessarily
equal to (13). The specific form of the
4-acceleration in (A2) implies the constraint

$$\left( {1/ L} \right)_{,t}={\textstyle{1 \over 3}}\Theta RJ(x^i)$$

\ni where $J$ is an arbitrary function. Inserting this constraint into (8),
and demanding the fulfilment of the Einstein field equations $G_{ij}=0$ and
$G_{ii}-G_{jj}=0$ ($i\ne j$), the metric (8) becomes (24a) with $J=f$.

\Par

\vskip .3cm
\ni {\bf References} \par
\vskip .3cm

\ni $^1$ D. Kramer, H. Stephani, M.A.H. MacCallum, and E. Herlt,
{\it Exact Solutions of Einstein's Field Equations},
(Cambridge University Press, Cambridge, UK, 1980). On the Stephani Universe,
see chapter 32, equation (32.46).

\Par

\ni $^2$ A. Barnes, {\it Gen. Rel. Grav.}, {\bf 4}, 105, (1973).
\Par

\ni $^3$ A. Krasinski, {\it Gen. Rel. Gravit.}, {\bf 15}, 673, (1983).
\Par

\ni $^4$ A. Krasinski, {\it J. Math. Phys.}, {\bf 30}, 433, (1989).
\Par

\ni $^5$ A. Krasinski,
{\it Physics in an Inhomomogeneous Universe, (a review)},
(Polish Research Committe and Department of Applied Mathematics, University
of Cape Town, South Africa). 1993.\Par

\ni $^6$ C. Bona and B. Coll, {\it C.R. Acad. Sci. Paris }, {\bf 1301},
(1985), 613.\Par

\ni $^7$ C. Bona and B. Coll, {\it Gen. Rel. Grav. }, {\bf 20},
(1988), 297.\Par

\ni $^8$  R.A. Sussman, {\it J. Math. Phys.}, {\bf 29}, 945, (1988). See
sections VI-VII. \Par

\ni $^9$ C.B. Collins, {\it {J. Math. Phys.}}, {\bf{26}}, 2009, (1985). See
also: C.B. Collins, {\it{Can. J. Phys.}}, {\bf{64}}, 191, (1986).
\Par

\ni $^{10}$ B. Coll and J.J. Ferrando, {\it {J. Math. Phys.} }, {\bf 26},
1583, (1985).\Par

\ni $^{11}$ H. Quevedo and R.A. Sussman, \qleft On the thermodynamics of
simple,
non-isentropic fluids in General Relativity\qright. Preprint submitted for
publication.\Par

\ni $^{12}$ R.A. Sussman, {\it Class. Quantum Grav.}, {\bf 11}, 1445,
(1994).\Par

\ni $^{13}$ Maple V (release 3), Waterloo Maple Software, 450 Philip St.,
Waterloo, Ontario, N2L 5J2 Canada.\Par

\ni $^{14}$ GRTensor (version 0.36), K. Lake and P. Musgrave,
Department of Physics, Queens University,
Kingston, Ontario, Canada.\Par

 \ni $^{15}$ A. Barnes and R.R. Rowlingson {\it Class. Quantum Grav.},
{\bf 7}, 1721, (1990).\Par

\ni $^{16}$ R. Maartens and S.D. Maharaj, {\it Class. Quantum Grav.}, {\bf 3},
1005, (1986).\Par

\ni $^{17}$ D.R. Oliver Jr and W.R. Davies, {\it Gen. Rel. Grav.},
{\bf 8}, 905, (1977). See also:  A.A. Coley, {\it Class. Quantum Grav.},
{\bf 8}, 955, (1991).\Par

\vfill\eject
\ni {\bf Figure Caption} \par
\vskip .3cm

\ni {\bf Figure 1. Thermodynamic quantities associated with the example
presented in section VI.}
\vskip .3cm

We chose the following values of the constant
free parameters: $a_1=10^{-4}$, $a_2=9\ \hbox{X}\ 10^{-5}$,
$a_3=1.1\ \hbox{X}\ 10^{-4}$,
$b_0=-9/20$, so that the positive solution of (32c) is $R_0\approx1.18$.
Figure (1a) displays $\Theta/3=R_{,t}/R$ vs. $R$, showing that the latter
is bounded between $R=0$ and $R=R_0$, with the latter constant being
the value of $R$ at which $\Theta$ vanishes. This means that $\rho$
evolves between $\rho=1$ at $R=R_0$ and infinity at $R=0$ (big bang and
big crunch singularities). Figure (1b) exhibits the ratio $p/\rho$
in terms of $R$ and the \qleft radial\qright coordinate $\chi$, for
$\theta=\pi/2$ and $\phi=0$. Since the constants $a_i$ are very small,
the plot looks qualitatively analogous for all values of these \qleft
angular\qright coordinates. Notice how the constraint $0<p/\rho\le 1$
holds throughout the evolution of the fluid, from $p/\rho=1$ at $R=0$,
keeping the same value along $\chi=0$ and evolving to $p/\rho\approx 0$
 at $R=R_0$ and $\chi\approx \pm\pi$. Figures (1c) and (1d) display
equations of state (31a) and (31c) The ratio $p/\rho$ and $T$ are plotted
in terms of $\arctan(\rho)$ in the range $(\arctan(1), \pi/2)$ for various
values of $S=S_0-\sigma$ in the range between $S=0$ and $S=S_0=9/20$.
Notice that $p/\rho\to1$ and $T\to\infty$ as $\arctan(\rho)\to\pi/2$
(that is: $\rho\to\infty$) for all values of $S$. As $\rho\to 1$, $T$
and $p/\rho$ decrease at various rates depending on the value of $S$.
The parameter values in these plots do not follow from
any physical consideration, we simply aim to illustrate that
free parameters exist so that the solutions satisfy the minimal physical
conditions discussed in section VI. These plots were obtained with the
symbolic computing program MAPLE.

\bye